\documentclass[aps,prl,twocolumn,showpacs]{revtex4}
\usepackage[dvips]{epsfig}
\begin{document}
\title{Long range intermolecular forces in triatomic systems:\\
connecting the atom-diatom and atom-atom-atom representations}
\author{Marko T. Cvita\v s, Pavel Sold\'{a}n and Jeremy M. Hutson}
\affiliation{Department of Chemistry, University of Durham, South
Road, Durham, DH1~3LE, England}

\date{\today}

\begin{abstract}
The long-range forces that act between three atoms are analysed in
both atom-diatom and atom-atom-atom representations. Expressions
for atom-diatom dispersion coefficients are obtained in terms of
3-body nonadditive coefficients. The anisotropy of atom-diatom
$C_6$ dispersion coefficients arises primarily from nonadditive
triple-dipole and quadruple-dipole forces, while pairwise-additive
forces and nonadditive triple-dipole and dipole-dipole-quadrupole
forces contribute significantly to atom-diatom $C_8$ coefficients.
The resulting expressions are applied to dispersion coefficients
for Li + Li$_2$ (triplet) and recommendations are made for the
best way to obtain global triatomic potentials that dissociate
correctly both to three separated atoms and to an atom and a
diatomic molecule.
\end{abstract}
\pacs{34.20.-b,34.20.Gj,34.20.Cf,34.30.+h} \maketitle

\font\smallfont=cmr7

\section{Introduction}

There is great current interest in forming diatomic molecules in
cold gases of alkali metal atoms, by photoassociation spectroscopy
\cite{Stw99,Wyn00,Ger00,McK02}, by magnetic tuning through
atom-atom Feshbach resonances
\cite{Mie00,Don02,Jin03b,Hul03,Sal03} and through 3-body
recombination \cite{Gri033body,Gri03BEC}. In most current
experiments, the diatomic molecules are formed in vibrational
states near dissociation and spend most of their time at large
internuclear separations. Once formed, the fate of the molecules
depends on collisional processes which in turn depend on atom-atom
and atom-diatom potential energy surfaces
\cite{Sol02,Que04,Cvi05Li3,Que05K3,Cvi05Limix,Cvi05Lipot}.

The potential energy surfaces for alkali metal trimers are
strongly nonadditive, even for spin-polarized atoms (quartet
electronic states) \cite{Hig00,Sol03}. In all cases both linear
and equilateral configurations of the M$_3$ collision complex lie
below the asymptotic atom-diatom energy, so that barrierless atom
exchange reactions can occur. We have carried out quantum dynamics
calculations including reactive channels for spin-polarized Na +
Na$_2$ \cite{Sol02}, Li + Li$_2$
\cite{Cvi05Li3,Cvi05Limix,Cvi05Lipot} and K + K$_2$
\cite{Que05K3}. In Na + Na$_2$, nonadditivity increases the well
depth by about 60\% \cite{Hig00} and increases the low-energy
cross sections for vibrational quenching by a factor of 10
\cite{Sol02}. For the other alkali metals the nonadditive
contributions to the potential are even larger \cite{Sol03}.

Low-energy collisions are particularly sensitive to long-range
potentials. In developing triatomic potential energy surfaces for
collision calculations, it is thus highly desirable to use global
functional forms that include nonadditivity and have the correct
physical behaviour both for three-body dissociation (to separated
atoms) and for two-body dissociation (to an atom and a diatomic
molecule). However, although there has been a considerable amount
of work on both these limits, the connection between them has not
been fully explored. This paper sets out to establish the
connection and to suggest functional forms that are correct in
both limits.

\section{Atom-diatom limit}

An atom-diatom system is conveniently described in terms of Jacobi
coordinates $R$, $r$ and $\theta$. In this case, ``long range" is
usually interpreted to mean that the atom-diatom distance $R$ is
large but that the diatom internal distance $r$ can be either
large or small. For an atom interacting with a homonuclear diatom
the long-range forces for $R\gg r$ (neglecting retardation and
damping) can be written to order $R^{-10}$ \cite{Buc67},
\begin{equation}
V(R,r,\theta) = -C_6(r,\theta) R^{-6} - C_8(r,\theta) R^{-8} -
C_{10}(r,\theta) R^{-10},
\end{equation}
where
\begin{eqnarray}
C_6(r,\theta) &=& C_6^0(r) + C_6^2(r) P_2(\cos\theta) \nonumber\\
C_8(r,\theta) &=&
C_8^0(r) + C_8^2(r) P_2(\cos\theta) + C_8^4(r) P_4(\cos\theta) \nonumber\\
C_{10}(r,\theta) &=& C_{10}^0(r) + C_{10}^2(r) P_2(\cos\theta)
\nonumber\\ &+& C_{10}^4(r) P_4(\cos\theta) + C_{10}^6(r)
P_6(\cos\theta). \label{eqatdiat}
\end{eqnarray}
The $r$-dependence of the dispersion coefficients $C_n^\lambda(r)$
is central to the present work.

R\'erat and Bussery-Honvault \cite{Rer03} have calculated
isotropic and anisotropic dispersion coefficients $C_6^0(r)$ and
$C_6^2(r)$ for Li and Na interacting with triplet Li$_2$ and
Na$_2$. They fitted the results to empirical functional forms
based on the known asymptotic behaviour of diatomic
polarizabilities \cite{Hei96}. However, they stated that the
proper asymptotic functional form for atom-diatom dispersion
coefficients was unknown. M\'erawa {\em et al.} \cite{Mer03}
extended parts of this work (not including the $r$-dependence of
the dispersion coefficients) to systems containing K and triplet
K$_2$.

\section{Separated atom limit}

A system of three well-separated atoms is more conveniently
described in terms of atom-atom distances $r_{12}$, $r_{23}$ and
$r_{31}$ and the angles $\phi_1$, $\phi_2$ and $\phi_3$ measured
at each atom. The interaction energy for such a system is
conventionally represented in terms of pairwise additive and
nonadditive terms,
\begin{equation}
\label{eqv3} V({\bf r}) = \sum_{i<j} V_{\rm dimer}(r_{ij}) +
V_{3}({\bf r}), \label{eqpair}
\end{equation}
where we use the shorthand $({\bf r})$ to indicate
$(r_{12},r_{23},r_{31},\phi_1,\phi_2,\phi_3)$. For a pair of
S-state atoms the long-range energy is
\begin{equation}
V_{\rm dimer}(r) = - C_6 r^{-6} - C_8 r^{-8} - C_{10} r^{-10} +
{\cal O}(r^{-11}).
\end{equation}
The nonadditive energy has several long-range contributions that
need careful consideration. The best-known is the
Axilrod-Teller-Muto (ATM) triple-dipole (DDD) term, which has the
form \cite{Axi43}
\begin{equation}
V_3^{\rm DDD}({\bf r}) = Z_{111} W_{111}({\bf r}),
\end{equation}
where \begin{equation} W_{111}({\bf r}) =
3(1+3\cos\phi_1\cos\phi_2\cos\phi_3) r_{12}^{-3} r_{23}^{-3}
r_{31}^{-3}. \label{eqW111}
\end{equation}
The triple-dipole term is one of several that arise in third-order
perturbation theory from terms of the type
\begin{equation} \frac{ \langle
000|H_{12}'|lm0\rangle \langle lm0|H_{23}'|l0n\rangle \langle
l0n|H_{31}'|000\rangle } {(\Delta E_l^1 + \Delta E_m^2) (\Delta
E_l^1 + \Delta E_n^3)}, \label{eqddd}
\end{equation}
where the ket $|lmn\rangle$ indicates a product wavefunction with
atoms 1, 2 and 3 in states $l$, $m$ and $n$ respectively and
$\Delta E_l^i$ is the excitation energy for state $l$ of atom $i$.
The interaction Hamiltonian $H_{ij}'$ is usually expanded at long
range in terms of multipole-multipole interactions. The
triple-dipole term arises when each of the three operators
$H_{ij}'$ is a dipole-dipole interaction of the form
\begin{equation}
H_{ij}'({\rm DD}) = \frac{\mu_i . \mu_j - 3(\mu_i.\hat r_{ij})
(\mu_j.\hat r_{ij})}{r_{ij}^3}, \end{equation} where $\hat r_{ij}$
is a unit vector along ${\bf r}_{ij}$. The triple-dipole term will
be referred to below as a third-order (3,3,3) contribution to
indicate the powers of the three distances involved.

There are additional third-order terms that arise when one or more
of the dipole operators is replaced by a quadrupole or
higher-order moment. The low-order terms and their resulting
contributions for 3 identical atoms are \cite{Bel70,Dor71}
\begin{eqnarray}
V_3^{\rm DDQ}({\bf r}) &=& Z_{112} W_{112}({\bf r})
\\
V_3^{\rm DQQ}({\bf r}) &=& Z_{122} W_{122}({\bf r})
\\
V_3^{\rm DDO}({\bf r}) &=& Z_{113} W_{113}({\bf r})
\\
V_3^{\rm QQQ}({\bf r}) &=& Z_{222} W_{222}({\bf r}) ,
\end{eqnarray}
where Q stands for quadrupole and O for octopole. The coefficients
$Z_{l_1l_2l_3}$ are related to polarizabilities of rank $l_1$,
$l_2$ and $l_3$ at imaginary frequencies,
\begin{equation} Z_{l_1l_2l_3} = (1/\pi)
\int_0^\infty \alpha^{l_1}(i\omega) \alpha^{l_2}(i\omega)
\alpha^{l_3}(i\omega) \,d\omega. \end{equation} The corresponding
geometric factors are
\begin{widetext}
\begin{eqnarray}
\label{eqW112} W_{112}({\bf r}) &=& \frac{3}{16}
[9\cos\phi_3-25\cos 3\phi_3 +
6\cos(\phi_1-\phi_2) (3+5\cos2\phi_3)] r_{12}^{-3} r_{23}^{-4} r_{31}^{-4} + {\rm c.p.}\\
W_{122}({\bf r}) &=& \frac{15}{64} [3(\cos\phi_1+5\cos5\phi_1) +
20\cos(\phi_2-\phi_3) (1-3\cos2\phi_1) \nonumber\\ &+&
70\cos2(\phi_2-\phi_3)\cos\phi_1]
 r_{12}^{-4} r_{23}^{-5} r_{31}^{-4} + {\rm c.p.}\\
W_{113}({\bf r}) &=& \frac{5}{32} [9 + 8\cos2\phi_3 - 49\cos4\phi_3 +
6 \cos(\phi_1-\phi_2)(9\cos\phi_3 + 7\cos3\phi_3)]
r_{12}^{-3} r_{23}^{-5} r_{31}^{-5} + {\rm c.p.}\nonumber\\ \\
\label{eqW222} W_{222}({\bf r}) &=& \frac{15}{128} \{-27 +
220\cos\phi_1\cos\phi_2\cos\phi_3 +
490\cos2\phi_1\cos2\phi_2\cos2\phi_3 \nonumber\\
&+& 175[\cos2(\phi_1-\phi_2) + \cos2(\phi_2-\phi_3) +
\cos2(\phi_3-\phi_1)]\} r_{12}^{-5} r_{23}^{-5} r_{31}^{-5},
\end{eqnarray}
where c.p.\ indicates summation of all cyclic permutations of
indices. It may be noted that the multipole operator on each atom
appears in {\it two} terms in Eq.\ (\ref{eqddd}), so that changing
one dipole operator into a higher-order one affects {\it two} of
the inverse powers.

In addition to these third-order terms, there are terms arising
from fourth and higher-order perturbation theory that make
important contributions to the long-range atom-diatom
coefficients. In principle, any combination of interaction
operators could produce a 4th-order term, although only those that
involve excitations on all three atoms produce a contribution to
the 3-body nonadditive energy. In addition, any odd-order
operators (dipoles and octopoles) must occur an {\it even} number
of times on each atom to satisfy parity constraints. Thus there
are terms such as
\begin{equation} \frac{ \langle
000|H_{12}'({\rm DD})|lm0\rangle \langle lm0|H_{23}'({\rm
DD})|lnp\rangle \langle lnp|H_{23}'({\rm DD})|lq0\rangle \langle
lq0|H_{12}'({\rm DD})|000\rangle } {(\Delta E_l^1 + \Delta E_m^2)
(\Delta E_l^1 + \Delta E_n^2 + \Delta E_p^3) (\Delta E_l^1 +
\Delta E_q^2)}, \label{eqdddd}
\end{equation}
which gives rise to a (6,6,0) contribution, and
\begin{equation} \frac{ \langle
000|H_{12}'({\rm DD})|lm0\rangle \langle lm0|H_{23}'({\rm
DD})|lnp\rangle \langle lnp|H_{12}'({\rm QQ})|q0p\rangle \langle
q0p|H_{31}'({\rm DD})|000\rangle } {(\Delta E_l^1 + \Delta E_m^2)
(\Delta E_l^1 + \Delta E_n^2 + \Delta E_p^3) (\Delta E_q^1 +
\Delta E_p^3)}, \label{eqDDDQ}
\end{equation}
which gives rise to a (8,3,3) contribution. However, terms such as
\begin{equation} \frac{ \langle
000|H_{12}'({\rm DD})|lm0\rangle \langle lm0|H_{23}'({\rm
DD})|lnp\rangle \langle lnp|H_{12}'({\rm DD})|q0p\rangle \langle
q0p|H_{31}'({\rm DD})|000\rangle } {(\Delta E_l^1 + \Delta E_m^2)
(\Delta E_l^1 + \Delta E_n^2 + \Delta E_p^3) (\Delta E_q^1 +
\Delta E_p^3)}
\end{equation}
\end{widetext}
are forbidden by parity, so that there is no (6,3,3) contribution.
The fourth-order dipole energy (\ref{eqdddd}) has been given
within a Drude oscillator model by Bade \cite{Bad57,Bad58},
\begin{equation}
V_3^{\rm DDDD}({\bf r}) = -\frac{45}{64} V \alpha^4
\left(1+\cos^2\phi_1\right)r_{12}^{-6} r_{13}^{-6} + {\rm c.p.},
\end{equation}
where $V$ is a characteristic excitation energy and $\alpha$ is
the static atomic dipole polarizability. If we adopt the
definition $Z_{1111}=(5/32)V\alpha^4$, which corresponds more
generally to
\begin{equation} Z_{l_1l_2l_3l_4} = (1/\pi)
\int_0^\infty \alpha^{l_1}(i\omega) \alpha^{l_2}(i\omega)
\alpha^{l_3}(i\omega) \alpha^{l_4}(i\omega) \,d\omega,
\end{equation}
then
\begin{equation}
V_3^{\rm DDDD}({\bf r}) = Z_{1111} W_{1111}({\bf r}),
\end{equation}
where
\begin{equation}
W_{1111}({\bf r}) =
-\frac{9}{2}\left(1+\cos^2\phi_1\right)r_{12}^{-6} r_{13}^{-6} +
{\rm c.p.} \label{eqW1111}
\end{equation}

In addition, there are fourth-order terms like Eq.\ (\ref{eqdddd})
but with one pair of the dipole-dipole operators replaced with
dipole-quadrupole operators. These give terms with powers
$(8,6,0)$, $(8,3,3)$, $(7,7,0)$, $(7,4,3)$ and $(6,4,4)$. The
angular factors have been given to within overall scaling factors
by Lotrich and Szalewicz \cite{Lot97} (though the $(6,3,3)$ term
that they describe is in fact zero and there are several
typographical errors in their equations). There is also a
fourth-order term involving a single octopole-dipole operator
(with all the rest dipole-dipole) that also contributes to
$(8,6,0)$ but is not mentioned in ref.\ \onlinecite{Lot97}.

There are also analogous terms arising from fifth-order
perturbation theory, which are also constrained by the requirement
that odd-order operators must occur an even number of times for
each atom. The leading such term (DDDDD) has powers (9,3,3), but
its explicit angular form has not been given previously. We have
evaluated it within the Drude model of ref. \onlinecite{Bad57}. We
adopt the definition $Z_{11111}=(35/256)V\alpha^5$, which
corresponds to
\begin{equation}
Z_{11111}= (1/\pi) \int_0^\infty \left[ \alpha^1(i\omega)\right]^5
\,d\omega.
\end{equation}
In the 5th-order case, Eq.\ (10) of ref. \onlinecite{Bad57} then
reduces to
\begin{equation} V_3^{\rm DDDDD}({\bf r})=Z_{11111} W_{11111}({\bf r}),
\end{equation} with
\begin{eqnarray} W_{11111}({\bf r}) &=&
9\left[2+3\cos^2\phi_1+3\cos^2\phi_2+\cos^2\phi_3\nonumber\right.\\
&+& \left. 9\cos\phi_1\cos\phi_2\cos\phi_3\right] r_{12}^{-9}
r_{23}^{-3} r_{31}^{-3} \nonumber\\&+& {\rm c.p.} \label{eqW11111}
\end{eqnarray}

\section{Connecting the atom-diatom and separated atom limits}

To make an explicit connection between the atom-atom additive and
nonadditive dispersion terms and the atom-diatom dispersion
formulae (\ref{eqatdiat}), we must collect terms in the
$(r_{12},r_{23},r_{31})$ representation that contribute to
individual powers of the Jacobi distance $R$. Let us consider the
case in which a diatom made up of atoms 1 and 2 is separated from
atom 3. For simplicity, we will assume that the three atoms are
identical. The atom-atom distances may be written in terms of
Jacobi coordinates $R$, $r$ and $\theta$,
\begin{eqnarray} r_{12} &=& r \\
r_{23} &=&
\sqrt{R^2+\frac{r^2}{4}+Rr\cos\theta}; \\
r_{31} &=& \sqrt{R^2+\frac{r^2}{4}-Rr\cos\theta},
\end{eqnarray}
and the cosines of $\phi_1$, $\phi_2$ and $\phi_3$ are given by
the cosine rule. The approach we take is to express the different
contributions to 3-body energies in Jacobi coordinates using these
equations and then to expand the results as power series in $r/R$.

We consider first the additive terms $V_{\rm dimer}(r_{ij})$. For
$R\gg r$ the atom-atom $C_6$ term makes contributions to the
atom-diatom coefficients of Eq.\ (\ref{eqatdiat}) given by
\begin{eqnarray}
C_6^{\rm add,6}(r,\theta) &=& 2C_6; \\
C_8^{\rm add,6}(r,\theta) &=& C_6 \left(\frac{5}{2} + 8P_2(\cos\theta)\right) r^2; \\
C_{10}^{\rm add,6}(r,\theta) &=& C_6 \left(\frac{7}{4} +
\frac{50}{7}P_2(\cos\theta) + \frac{48}{7}P_4(\cos\theta)\right)
r^4.\nonumber\\
\end{eqnarray}
Similarly, the atom-atom $C_8$ and $C_{10}$ terms contribute
\begin{eqnarray}
C_8^{\rm add,8}(r,\theta) &=& 2C_8; \\
C_{10}^{\rm add,8}(r,\theta) &=& C_8 \left(\frac{14}{3} + \frac{40}{3}P_2(\cos\theta)\right) r^2;\\
C_{10}^{\rm add,10}(r,\theta) &=& 2C_{10}.
\end{eqnarray}
It may be seen that the atom-atom pair potential contributes {\it
no} anisotropy to the longest-range ($R^{-6}$) term in the
atom-diatom potential. All such anisotropy must come from 3-body
nonadditive terms in the potential. The only third-order 3-body
term that contributes to $C_6(r,\theta)$ is the triple-dipole
term. Its geometric factor may be expanded at large $R$,
\begin{widetext}
\begin{equation}
W_{111}(R,r,\theta) = -6P_2(\cos\theta) r^{-3}R^{-6} +
\left(\frac{3}{2} - \frac{6}{7}P_2(\cos\theta) -
\frac{36}{7}P_4(\cos\theta)\right) r^{-1}R^{-8} + {\cal
O}(rR^{-10}).
\end{equation}
Similarly, the geometric factors for the third-order DDQ, DQQ and
DDO terms may be expanded
\begin{eqnarray}
W_{112}(R,r,\theta) &=& -\frac{120}{7} \left[P_2(\cos\theta) -
P_4(\cos\theta) \right] r^{-3}R^{-8} + {\cal
O}(r^{-1}R^{-10}) \\
W_{122}(R,r,\theta) &=& 30 P_4(\cos\theta) r^{-5}R^{-8} + {\cal
O}(r^{-3}R^{-10}) \\
W_{113}(R,r,\theta) &=& -40 P_4(\cos\theta) r^{-5}R^{-8} + {\cal
O}(r^{-3}R^{-10}),
\end{eqnarray}
where any contributions from cyclic permutations are now included.
These three terms thus contribute to the atom-diatom $C_8(r)$ and
its anisotropy but not to $C_6(r,\theta)$.

As noted above, the fourth-order DDDD term does not have a (6,3,3)
contribution but does have a (6,6,0) contribution. It can thus
contribute to the atom-diatom $C_6$ coefficient. Its geometric
factor (\ref{eqW1111}) may be expanded at large $R$,
\begin{eqnarray}
W_{1111}(R,r,\theta) = &-&6 \left[2+P_2(\cos\theta)\right]
r^{-6}R^{-6} \nonumber\\ &-&
\left(\frac{33}{2}+\frac{402}{7}P_2(\cos\theta)+\frac{144}{7}P_4(\cos\theta)\right)
r^{-4}R^{-8}  + {\cal O}(r^{-2}R^{-10}). \label{eqW1111jac}
\end{eqnarray}
Collecting these equations together provides expressions for the
behaviour of the atom-diatom dispersion coefficients
$C_n^\lambda(r)$ at large $r$,
\begin{eqnarray}
\label{c60}
 C_6^0(r) &=& 2C_6 + 12 Z_{1111} r^{-6} +
{\cal O}(r^{-8}); \\
\label{c62} C_6^2(r) &=& 6Z_{111} r^{-3} + 6 Z_{1111}
r^{-6} + {\cal O}(r^{-8}); \\
\label{c80} C_8^0(r) &=& \frac{5}{2}C_6r^2 + 2C_8 -\frac{3}{2}
Z_{111} r^{-1} + \frac{33}{2} Z_{1111}r^{-4}
+ {\cal O}(r^{-6}); \\
C_8^2(r) &=& 8C_6r^2 + \frac{6}{7} Z_{111} r^{-1} + \frac{120}{7}
Z_{112} r^{-3} +\frac{402}{7}Z_{1111}r^{-4}+ {\cal O}(r^{-6}); \\
\label{c84} C_8^4(r) &=& \frac{36}{7} Z_{111} r^{-1} -
\frac{120}{7} Z_{112} r^{-3} + \frac{144}{7}Z_{1111}r^{-4} +
\left( 40 Z_{113} - 30 Z_{122}\right) r^{-5} + {\cal O}(r^{-6}).
\end{eqnarray}
\end{widetext}
The ${\cal O}(r^{-8})$ terms in Eqs.\ (\ref{c60}) and (\ref{c62})
come from the $(8,6,0)$, $(8,3,3)$ and $(6,4,4)$ contributions to
the fourth-order energy, while the ${\cal O}(r^{-6})$ terms in
Eqs.\ (\ref{c80}) to (\ref{c84}) come from both these and the
$(7,7,0)$ and $(7,4,3)$ contributions. The fifth-order term (Eq.\
(\ref{eqW11111})) does not contribute until ${\cal
O}(r^{-9}R^{-6})$.

\section{Fitting dispersion coefficients for L\lowercase{i} +
L\lowercase{i}$_2$}

The equations above apply when all three of $r_1$, $r_2$ and $r_3$
are large. However, when any of them is small, the power series is
insufficient. We therefore use Eqs.\ (\ref{c60}) to (\ref{c84}) as
the long-range limits of more general expressions, constructed by
(i) multiplying the individual inverse power terms by damping
functions $D_n(r)$ and (ii) adding a short-range exponential term
to allow for the effects of orbital overlap.

As described above, R\'erat and Bussery-Honvault \cite{Rer03} have
calculated isotropic and anisotropic dispersion coefficients for
Li + Li$_2$ $(^3\Sigma_u^+)$ as a function of $r$ and have fitted
them to long-range expansions. They stated that ``no asymptotic
form of the coefficients $C_6$ exists to our knowledge", but found
empirically that $C_6^2(r)$ required both $r^{-3}$ and $r^{-6}$
terms, while $C_6^0(r)$ required only $r^{-6}$. Our expressions
(\ref{c60}) and (\ref{c62}) above provide the explanation for
this. However, our results also show that there should be
relationships among the coefficients of the fit, and the resulting
constraints were not included in ref.\ \onlinecite{Rer03}. In
particular, the coefficient of $r^{-6}$ in $C_6^2$ should be half
that in $C_6^0$ and $Z_{1111}$ can be related at least
approximately to $Z_{111}$ and $C_6$ as described below.

In devising functional forms for $C_6^0(r)$ and $C_6^2(r)$ it is
important to consider damping of the inverse power terms. For the
two-body interaction energy, the most popular approach is to use
Tang-Toennies damping functions \cite{Tan84} of the form
\begin{equation}
D_n(R) = 1 - e^{-bR} \sum_{k=0}^n \frac{(bR)^k}{k!}.
\end{equation}
When damping 3-body terms such as Eqs.\ (\ref{eqW111}) and
(\ref{eqW112}) to (\ref{eqW222}), a damping function is required
for each $r_i^n$. We have chosen to use $\sqrt{D_{2n}(r_i)}$
rather than $D_n(r_i)$ for this purpose, because this recovers the
correct $D_n(R)$ in the 2-body energies. Thus we damp 3-body terms
according to prescriptions such as
\begin{eqnarray}
\label{eqdamp333}  r_{12}^{-3} r_{23}^{-3} r_{31}^{-3} \rightarrow
r_{12}^{-3} r_{23}^{-3} r_{31}^{-3} \sqrt{D_6(r_{12})
D_6(r_{23}) D_6(r_{31})};\\
\label{eqdamp344}  r_{12}^{-3} r_{23}^{-4} r_{31}^{-4} \rightarrow
r_{12}^{-3} r_{23}^{-4} r_{31}^{-4} \sqrt{D_6(r_{12}) D_8(r_{23})
D_8(r_{31})}.
\end{eqnarray}
When damping is introduced in this way, the $r^{-6}$ terms in the
expressions for dispersion coefficients are damped with $D_6(r)$,
but the $r^{-3}$ term in $C_6^2(r)$ is damped with $\sqrt{D_6(r)}$
rather than $D_3(r)$. The expressions that we fit to are therefore
\begin{widetext}
\begin{eqnarray}
\label{eqfitc60} C_6^0(r) &=& 2C_6 + 12 Z_{1111} r^{-6} D_6(r)
+ A\exp(-Cx); \\
\label{eqfitc62} C_6^2(r) &=& 6Z_{111} r^{-3} \sqrt{D_6(r)} + 6
Z_{1111} r^{-6} D_6(r) + B\exp(-Cx),
\end{eqnarray}
\end{widetext} where $x=(r-r_0)/r_0$ and $r_0=7.0$ \AA.

\begin{figure} [htbp]
\begin{center}
\rotatebox{270}{ \resizebox{6.9cm}{!} {\includegraphics{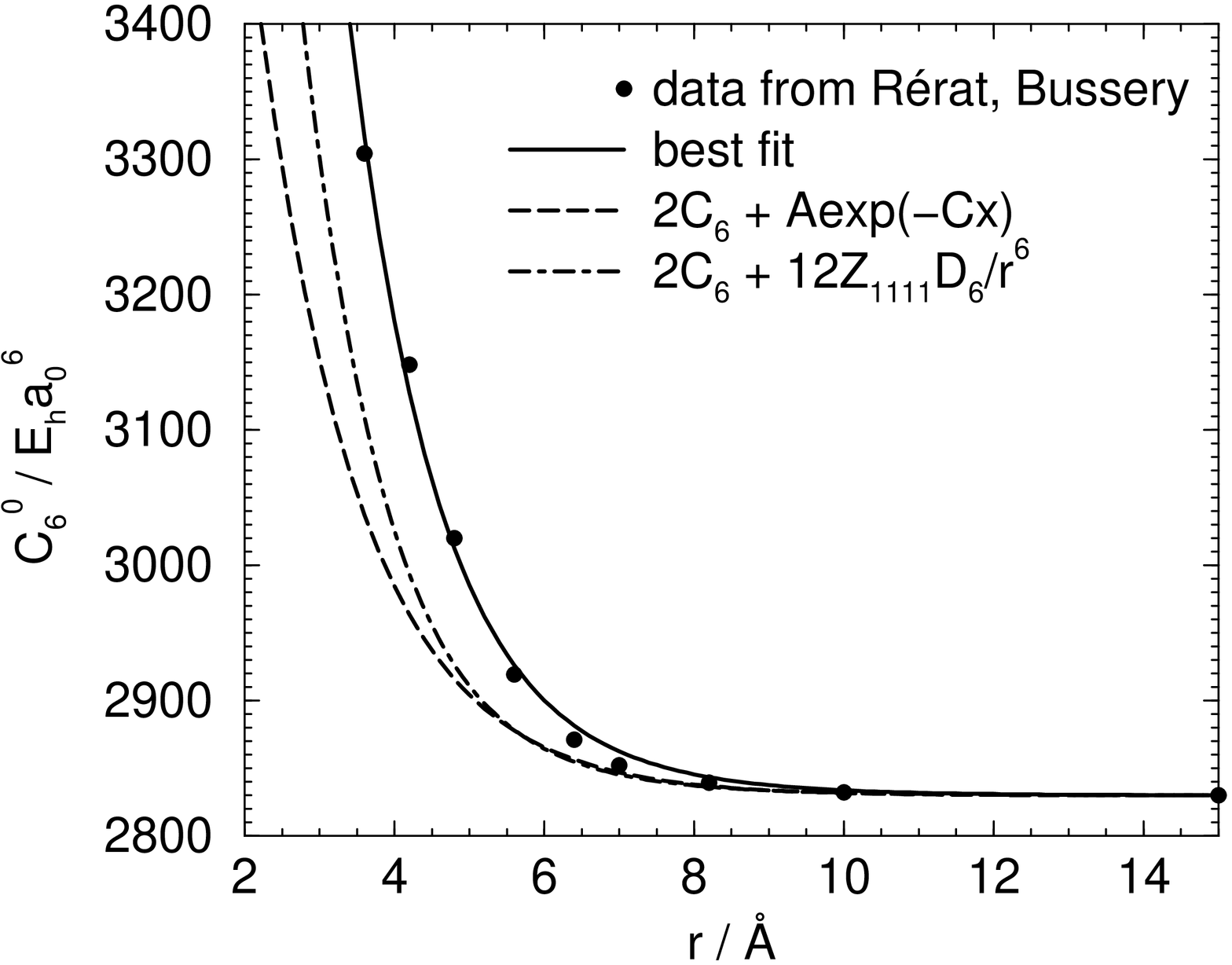}}}
\rotatebox{270}{ \resizebox{6.9cm}{!} {\includegraphics{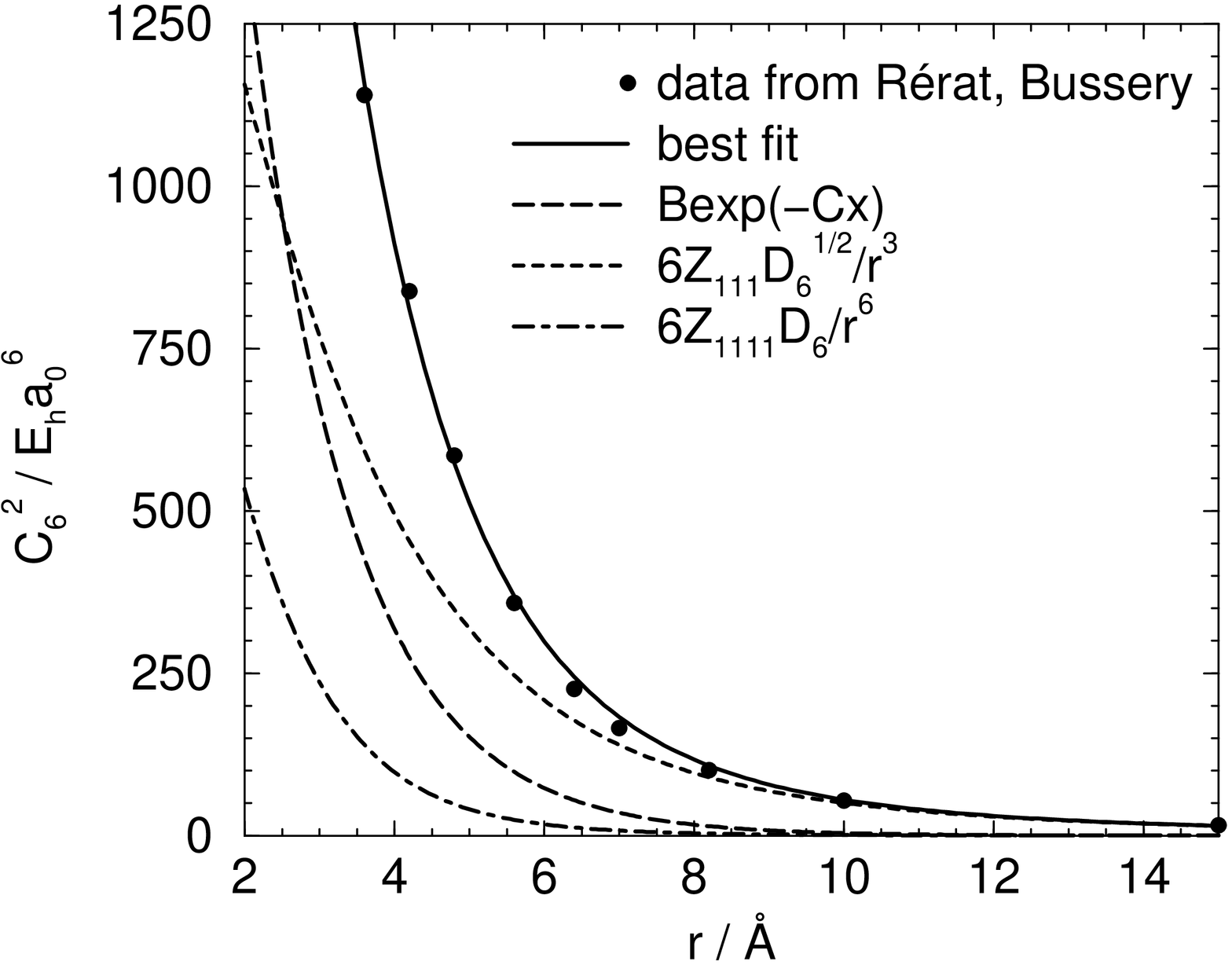}}}
\caption{Fits to dispersion coefficients $C_6^0(r)$ (upper panel)
and $C_6^2(r)$ (lower panel) for Li + Li$_2$
interactions.}\label{c6fig}
\end{center}
\end{figure}
We have fitted the values of $C_6^0(r)$ and $C_6^2(r)$ in ref.\
\onlinecite{Rer03} to the functional forms (\ref{eqfitc60}) and
(\ref{eqfitc62}). [Our $C_n^\lambda(r)$ are
$C_n^{\lambda0\lambda}(r)/\sqrt{\lambda(\lambda+1)}$ in the
notation of ref.\ \onlinecite{Rer03}.] The triple-dipole
coefficient was fixed at a value taken from variational
calculations with Hylleraas basis sets by Yan {\em et al.}
\cite{Yan96}, $Z_{111} = \nu_{abc}/3 = 5.687\times10^4\ E_ha_0^9$
(where $a_0$ is the Bohr radius and $E_h$ is the Hartree energy).
There is no {\em ab initio} value available for the
quadruple-dipole coefficient, but in a Drude model
$Z_{1111}=(5/32)V\alpha^4$ as described above, where $V$ is a
characteristic excitation energy and $\alpha$ is the atomic dipole
polarizability. Within the same model, $C_6=(3/4)V\alpha^2$ and
$Z_{111}=(3/16)V\alpha^3$. Combining these results gives an
estimate \begin{equation} Z_{1111} \approx
\frac{10(Z_{111})^2}{3C_6}. \end{equation} For Li$_3$ with $C_6$
and $Z_{111}$ values from ref.\ \onlinecite{Yan96}, this gives
$Z_{1111}=7.73\times10^6\ E_ha_0^{12}$. The $C_6$ coefficient in
Eq.\ (\ref{eqfitc60}) could not be fixed at the value $C_6=1393\
E_ha_0^6$ from ref.\ \onlinecite{Yan96}, because the results for
$C_6(r)$ in ref.\ \onlinecite{Rer03} converge on a slightly
different asymptotic value, so $C_6$ was allowed to vary in our
fit.

Meath and coworkers \cite{Kno86,Whe93} have calculated dispersion
damping functions for Li$_2$ and found that the $r^{-6}$ damping
function is around 0.45 at the diatomic minimum, $r_e\approx4.2$
\AA. With the Tang-Toennies form of $D_6(r)$, this requires
$b\approx1.5$ \AA$^{-1}$. We found that the exponent $C$ needed to
fit the short-range part of $C_6(r)$ was very different from
$br_0$, so we allowed $br_0$ and $C$ to be different and fixed $b$
at 1.5 \AA$^{-1}$. The remaining parameters were then determined
by a weighted least-squares fit, giving values $C_6=1414.8\
E_ha_0^6$, $A=17.2\pm2.6\ E_ha_0^6$, $B=35.2\pm3.9\ E_ha_0^6$ and
$C=5.13\pm0.25$. The quality of the resulting fit is shown in
Fig.\ \ref{c6fig}, together with the contributions of the
individual terms in Eqs.\ (\ref{eqfitc60}) and (\ref{eqfitc62}).

R\'erat and Bussery-Honvault \cite{Rer03} have also given values
of $C_8^0(r)$, $C_8^2(r)$ and $C_8^4(r)$ for the single distance
$r=4.2$~\AA, which is near the equilibrium distance for triplet
Li$_2$. In our notation their values correspond to $3.92 \times
10^5\ E_ha_0^8$, $1.91 \times 10^5\ E_ha_0^8$ and $-0.38 \times
10^5\ E_ha_0^8$, respectively, where $E_h$ is the Hartree energy
and $a_0$ is the Bohr radius. Evaluating the long-range
expressions (\ref{c80}) to (\ref{c84}) at $r=4.2$ \AA, using
$C_6=1393\ E_ha_0^6$ and $Z_{111}=5.687 \times 10^4\ E_ha_0^9$ as
above with $C_8=83426\ E_ha_0^8$ \cite{Yan96},
$Z_{112}=5.81\times10^5\ E_ha_0^{11}$, $Z_{122}=6.41\times10^6\
E_ha_0^{13}$ and $Z_{113}=1.70\times10^7\ E_ha_0^{13}$
\cite{Pat97}, gives $4.076 \times 10^5\ E_ha_0^8$, $8.402 \times
10^5\ E_ha_0^8$ and $+0.725 \times 10^5\ E_ha_0^8$, respectively.
It thus appears that $r=4.2$ \AA\ is too small a distance for the
$C_8(r)$ coefficients to be dominated by undamped long-range
contributions, and indeed it may be seen from Fig.\ \ref{c6fig}
that the exponential contributions to $C_6^0(r)$ and $C_6^2(r)$
are quite large at this distance. It would be very interesting to
calculate $r$-dependent $C_8$ coefficients and explore the onset
of long-range behaviour.

\section{Global functional forms for potential energy surfaces}

The results obtained above have important implications for the
choice of functional forms to represent potential energy surfaces
for triatomic systems. For low-energy scattering calculations, it
is highly desirable to have an interaction potential that
dissociates correctly both to three atoms and to an atom and a
diatomic molecule. This is especially important for processes such
as 3-body recombination and collisions of long-range diatomic
molecules with atoms, which are of current interest in studies of
cold molecule formation
\cite{Mie00,Don02,Jin03b,Hul03,Sal03,Gri033body,
Gri03BEC,Sol02,Que04,Cvi05Li3,Que05K3,Cvi05Limix,Cvi05Lipot}.

A global potential energy surface for a homonuclear triatomic
system must be symmetric in the atom indices if it is to reproduce
the full symmetry of the system. The simplest coordinate system
that achieves this is a set of 3 bond length coordinates
$(r_{12},r_{23},r_{31})$. Jacobi coordinates cannot easily
describe the full symmetry.

Our approach is to obtain a potential energy surface from
high-quality {\em ab initio} calculations on a grid of values
$(r_{12},r_{23},r_{31})$. Once this has been done, we need to
interpolate and extrapolate it in a way that incorporates the
correct long-range behaviour. In one dimension, reciprocal power -
reproducing kernel Hilbert space (RP-RKHS) interpolation
\cite{Ho96,Ho00,Sol00} provides an attractive way to obtain a
potential with desired inverse power behaviour at long range.
However, multidimensional RP-RKHS interpolation \cite{Ho96,Hig00}
at best gives a potential that extrapolates beyond the points as a
simple product of inverse powers in the different coordinates.
None of the long-range 3-body terms have this form, so a more
sophisticated approach is required.

Two different cases may be distinguished. For systems such as
spin-polarized Na$_3$ or K$_3$, the nonadditive forces are
substantial at short range but do not dwarf the additive forces
\cite{Hig00,Sol03}. Under these circumstances, we have found it
best to transform the potential to a form that does behave as a
simple product of inverse powers at long range and then
interpolate in that form. The first stage in this process is to
subtract the (assumed known) sum of pair potentials from the {\it
ab initio} points according to Eq.\ (\ref{eqv3}) to obtain the
nonadditive contribution to the interaction, $V_3({\bf r})$. The
leading terms in the long-range expansion of $V_3({\bf r})$ are
the DDD and DDQ terms. However, it may be noted that the DDD term
of Eq.\ (\ref{eqW111}) vanishes on a seam in the angular space and
the DDQ term of Eq.\ (\ref{eqW112}) vanishes at all linear
configurations. It is therefore not adequate to divide $V_3({\bf
r})$ by an angular factor in order to obtain a form that depends
only on inverse powers. Fortunately, the coefficients $Z_{111}$
and $Z_{112}$ are known for all the homonuclear alkali metal atom
systems \cite{Pat97}, so that damped versions of these terms can
be subtracted from the total nonadditive energy $V_3$ to give a
quantity $V'_3$,
\begin{equation} V_3'({\bf r}) = V_3({\bf r}) -
\left[V^{\rm DDD}_{\rm 3,damp}({\bf r}) + V^{\rm DDQ}_{\rm
3,damp}({\bf r}) \right].\end{equation} In our work on K$_3$
\cite{Que05K3}, we used a single damping function for both terms.
However, the present work has shown that it would be better to
choose separate damping functions for each inverse power term as
in Eqs.\ (\ref{eqdamp333}) and (\ref{eqdamp344}), and to define
(for example) \begin{eqnarray} V^{\rm DDQ}_{\rm 3,damp}({\bf r})
&=& \frac{3}{16} Z_{112} [9\cos\phi_3-25\cos 3\phi_3 \nonumber\\
&+&
6\cos(\phi_1-\phi_2) (3+5\cos2\phi_3)] \nonumber\\
&\times& \frac{\sqrt{D_6(r_{12}) D_8(r_{23}) D_8(r_{31})}}
{r_{12}^{3} r_{23}^{4} r_{31}^{4}} + {\rm c.p.}
\end{eqnarray}
The leading term in the asymptotic multipole expansion of $V_{3}'$
is the fourth-order dipole-dipole-dipole term (DDDD), which has
the more complicated (unfactorizable) form of Eq.\ (\ref{eqW1111})
above, with powers (6,6,0). If $Z_{1111}$ is known, this too could
be subtracted out. However, this term is negative at all
geometries, so a satisfactory alternative is to eliminate it by
defining $V_{3}''({\bf r}) = g({\bf r}) \times V_{3}'({\bf r})$,
where
\begin{widetext}
\begin{equation}
g({\bf r}) =
\frac{r_{12}^3r_{23}^3r_{13}^3}{(1+\cos^2\phi_{1})\,r_{23}^{6}+
(1+\cos^2\phi_{2})\,r_{13}^{6}+(1+\cos^2\phi_{3})\,r_{12}^6}.
\label{V3pp}
\end{equation}
\end{widetext}
The leading asymptotic term of the function $V_{3}''$ now has the
form $-\hbox{constant}\times r_{12}^{-3}r_{23}^{-3}r_{13}^{-3}$
and is suitable for an ``isotropic'' extrapolation of the type
that results from a multidimensional RP-RKHS interpolation.

The approach that we use is therefore to construct $V_{3}''$ at
the {\it ab initio} points as above and then interpolate it using
the fully symmetrized 3D RP-RKHS interpolation method
\cite{Hig00}. An RP-RKHS interpolation with respect to $r^p$ with
RKHS parameters $n$ and $m$ gives a potential with leading
long-range powers $r^{-p(m+1)}$ and $r^{-p(m+2)}$. Appropriate
choices thus include $p=3,m=0$ (as in our work on K$_3$
\cite{Que05K3}) and $p=1,m=2$. The interpolated potential is then
rebuilt as
\begin{equation}
V_3({\bf r}) = \frac{V_3''({\bf r})}{g({\bf r})} + \left[V^{\rm
DDD}_{\rm 3,damp}({\bf r}) + V^{\rm DDQ}_{\rm 3,damp}({\bf
r})\right]. \label{Vorig}
\end{equation}

A system such as quartet Li$_3$ requires a different approach
\cite{Cvi05Lipot}. In this case the nonadditive forces are so
large that at short range it does not make sense to decompose the
potential into additive and nonadditive parts at all. The
nonadditive potential is many times larger than the additive
potential \cite{Sol03}, and the decomposition would require the
final short-range potential to be expressed as the difference of
two large numbers. Nevertheless, at long range a decomposition
according to Eq.\ (\ref{eqv3}) is essential. Under these
circumstances, we found it best to carry out an unconstrained fit
to the {\it ab initio} points at short range, without imposing the
correct long-range behaviour, to obtain a short-range function
$V_{\rm SR}({\bf r})$. We then use a switching function $S({\bf
r})$ to join this onto the correct long-range form. We thus have
\begin{equation}
V({\bf r}) = S({\bf r}) V_{\rm SR}({\bf r}) + [1-S({\bf r})]
V_{\rm LR}({\bf r}).
\end{equation}
The long-range form must be valid when {\it any} of the atom-atom
distances is large. In our work on Li$_3$ \cite{Cvi05Lipot}, we
used
\begin{equation} V_{\rm LR}({\bf r}) = \sum_{i<j} V_{\rm
dimer}(r_{ij}) + V_{\rm 3,LR}({\bf r}),
\end{equation}
where
\begin{equation} V_{\rm 3,LR}({\bf r})= V_{\rm 3,damp}^{\rm DDD}({\bf r})
+ V_{\rm 3,damp}^{\rm DDQ}({\bf r}) + V_{\rm 3,damp}^{\rm
DDDD}({\bf r}) + V_{\rm 3,rep}^{\rm DD}({\bf r}).
\end{equation}
The function of the term $V_{\rm 3,rep}^{\rm DD}({\bf r})$ is to
ensure that the atom-diatom dispersion coefficients $C_6^0(r)$ and
$C_6^2(r)$ have the correct values (given by Eqs.\
(\ref{eqfitc60}) and (\ref{eqfitc62})) even when one of the
atom-atom distances is small. When $R\gg r$, this is achieved by
defining
\begin{eqnarray} \label{eqexch}
V_{\rm 3,rep}^{\rm DD}({\bf r}) &=& -\left[A + B
P_2(\cos\theta)\right] \nonumber\\
&\times& \exp(-Cx)\, r_{23}^{-3} r_{31}^{-3} \sqrt{D_6(r_{23})
D_6(r_{31})} \nonumber\\&+& {\rm c.p.},
\end{eqnarray}
where $x=(r_{12}-r_0)/r_0$ as before and the parameters $A$, $B$
and $C$ come from fits to numerical values of $C_6^0(r)$ and
$C_6^2(r)$ as described above. In evaluating Eq.\ (\ref{eqexch})
it is convenient to use an approximate form of $P_2(\cos\theta)$
that is valid for $R\gg r$ but is well-behaved at all geometries,
\begin{equation}
P_2(\cos\theta) \approx -\frac{1}{2} \left( 1 + 3 \cos\phi_1
\cos\phi_2 \cos\phi_3\right). \end{equation} This is already
evaluated as part of $W_{111}({\bf r})$, and since it is symmetric
the cyclic permutations required in Eq.\ (\ref{eqexch}) involve no
extra geometric calculations.

Finally, the switching function $S({\bf r})$ must become zero when
{\it any} of the three atom-atom distances is large. For Li$_3$ we
chose to use
\begin{equation}
S({\bf r}) = \frac{1}{2} \tanh [1-s_1(r_1+r_2+r_3-s_2)],
\end{equation}
with the parameters $s_1$ and $s_2$ determined in such a way that
the switching takes place in a region where both functional forms
give reasonably accurate energies.

\section{Conclusions}
We have investigated the relationship between long-range
intermolecular forces for triatomic systems in the atom-diatom and
atom-atom-atom representations. We have obtained expressions
relating the dispersion coefficients in the two representations.
We have shown that the anisotropy of the atom-diatom $C_6$
dispersion coefficient arises entirely from nonadditive terms in
the 3-body expansion. The most significant contributions at long
range arise from the third-order triple-dipole term and the
fourth-order quadruple-dipole term. The leading contributions to
the atom-diatom $C_8$ coefficient arise from the additive
atom-atom $C_6$ and $C_8$ coefficients and the third-order
nonadditive triple-dipole and dipole-dipole-quadrupole
coefficients.

There is great current interest in the formation of diatomic
molecules in cold atomic gases, and the collisional properties of
such molecules are of great importance. Calculations on these
collisions need triatomic interaction potentials that dissociate
properly both to an atom and a diatomic molecule and to three
separated atoms. We have used our results to suggest strategies
for obtaining such potentials.

\section{acknowledgments}

PS and JMH are grateful to EPSRC for support under research grant
GR/R17522/01. MTC is grateful for sponsorship from the University
of Durham and Universities UK.


\begin{thebibliography}{10}
\goodbreak
\bibitem{Stw99}
W. C. Stwalley and H. Wang,
{J. Mol. Spectrosc.} {\bf 195}, 194 (1999).

\bibitem{Wyn00}
R. Wynar, R. S. Freeland, D. J. Han, C. Ryu, and D. J. Heinzen,
{Science} {\bf 287}, 1016 (2000).

\bibitem{Ger00}
J. M. Gerton, D. Strekalov, I. Prodan and R. G. Hulet,
{Nature} {\bf 408}, 692 (2000).

\bibitem{McK02}
C. McKenzie, J. H. Denschlag, H. H\"affner, A. Browaeys, {\it et
al.}, {Phys. Rev. Lett.} {\bf 88}, 120403 (2002).

\bibitem{Mie00}
F. H. Mies, E. Tiesinga and P. S. Julienne, {Phys.\ Rev.\ A} {\bf
61}, 022721 (2000).

\bibitem{Don02}
E. A. Donley, N. R. Claussen, S. T. Thompson and C. E. Wieman,
Nature {\bf 417}, 529 (2002).

\bibitem{Jin03b}
C. A. Regal, C. Ticknor, J. L. Bohn, and D. S. Jin, {Nature
(London)} \textbf{424}, 47 (2003).

\bibitem{Hul03}
K. E. Strecker, G. B. Partridge and R. G. Hulet, {Phys. Rev.
Lett.} \textbf{91}, 080406 (2003).

\bibitem{Sal03}
J. Cubizolles, T. Bourdel, S. J. J. M. F. Kokkelmans, G. V.
Shlyapnikov, and C. Salomon, {Phys. Rev. Lett.} \textbf{91},
240401 (2003).

\bibitem{Gri033body}
S. Jochim, M. Bartenstein, A. Altmeyer, G. Hendl, C. Chin, J.
Hecker Denschlag, and R. Grimm, {Phys. Rev. Lett.} \textbf{91},
240402 (2003).

\bibitem{Gri03BEC}
S. Jochim, M. Bartenstein, A. Altmeyer, G. Hendl, S. Riedl, C.
Chin, J. Hecker Denschlag, and R. Grimm, {Science} \textbf{302},
2101 (2003).

\bibitem{Sol02}
P. Sold\'{a}n, M. T. Cvita\v{s}, J. M. Hutson, P. Honvault, and
J.-M. Launay, {Phys.\ Rev.\ Lett.} {\bf 89}, 153201 (2002).

\bibitem{Que04}
G. Qu\'{e}m\'{e}ner, P. Honvault, and J.-M. Launay, {Eur.\ Phys.\
J. D} {\bf 30}, 201 (2004).

\bibitem{Cvi05Li3}
M. T. Cvita\v{s}, P. Sold\'{a}n, J. M. Hutson, P. Honvault, and
J.-M. Launay, {Phys. Rev. Lett.} \textbf{94}, 033201 (2005).

\bibitem{Que05K3}
G. Qu\'{e}m\'{e}ner, P. Honvault, J.-M. Launay, P. Sold\'{a}n,
D.~E. Potter and J. M. Hutson, {Phys. Rev. A} {\bf 71}, 032722
(2005).

\bibitem{Cvi05Limix}
M. T. Cvita\v{s}, P. Sold\'{a}n, J. M. Hutson, P. Honvault, and
J.-M. Launay, {Phys. Rev. Lett.}, in press for issue of 3 June
2005. Preprint available from
http://arxiv.org/abs/cond-mat/0501636.

\bibitem{Cvi05Lipot}
M. T. Cvita\v{s}, P. Sold\'{a}n, J. M. Hutson, P. Honvault, and
J.-M. Launay, to be published.

\bibitem{Hig00}
J.~Higgins, T. Hollebeek, J. Reho, T.-S. Ho, K. K. Lehmann, H.
Rabitz, and G. Scoles, {J.~Chem. Phys.} {\bf 112}, 5751 (2000).

\bibitem{Sol03}
P. Sold\'{a}n, M. T. Cvita\v{s}, and J. M. Hutson, {Phys.\ Rev.\
A} {\bf 67}, 054702 (2003).

\bibitem{Buc67}
A. D. Buckingham, {Adv. Chem. Phys.} {\bf 12}, 107 (1967)

\bibitem{Rer03}
M. R\'erat and B. Bussery-Honvault, {Molec. Phys.} {\bf 101}, 373
(2003).

\bibitem{Hei96}
T. G. A. Heijmen, R. Moszynski, P. E. S. Wormer and A. van der
Avoird, {Molec. Phys.} {\bf 89}, 81 (1996).

\bibitem{Mer03}
M. M\'erawa, M. R\'erat and B. Bussery-Honvault, {J. Mol. Struct.
(Theochem)} {\bf 633}, 137 (2003).

\bibitem{Axi43}
B. M. Axilrod and E. Teller, {J. Chem. Phys.} {\bf 11}, 299
(1943).

\bibitem{Bel70}
R. J. Bell, {J. Phys. B} {\bf 3}, 751 (1970).

\bibitem{Dor71}
M. B. Doran and I. J. Zucker, {J. Phys. C} {\bf 4}, 307 (1971).

\bibitem{Bad57}
W. L. Bade, {J. Chem. Phys.} {\bf 27}, 1280 (1957).

\bibitem{Bad58}
W. L. Bade, {J. Chem. Phys.} {\bf 28}, 282 (1958).

\bibitem{Lot97}
V. F. Lotrich and K. Szalewicz, {J. Chem. Phys.} {\bf 106}, 9688
(1997).

\bibitem{Tan84}
K. T. Tang and J. P. Toennies, {J. Chem. Phys.} {\bf 80}, 3726
(1984).

\bibitem{Yan96}
Z.-C. Yan, J. F. Babb, A. Dalgarno and G. W. F. Drake, {Phys. Rev.
A} {\bf 54}, 2824 (1996).

\bibitem{Kno86}
P. J. Knowles and W. J. Meath, {Chem. Phys. Lett.} {\bf 124}, 164
(1986).

\bibitem{Whe93}
R. J. Wheatley and W. J. Meath, {Molec. Phys.} {\bf 80}, 25
(1993).

\bibitem{Pat97}
S. H. Patil and K. T. Tang, {J. Chem. Phys.} {\bf 106}, 2298
(1997).


\bibitem{Ho96}
T-S. Ho and H. Rabitz, {J. Chem. Phys.} \textbf{104}, 2584 (1996).

\bibitem{Ho00}
T-S. Ho and H. Rabitz, {J. Chem. Phys.} \textbf{113}, 3960 (2000).

\bibitem{Sol00}
P. Sold\'{a}n and J. M. Hutson, {J. Chem. Phys.} \textbf{112},
4415 (2000).


\end{thebibliography}
\end{document}